# Identifying synthetic voices' qualities for conversational agents

M. Cuciniello[1], T. Amorese[1], G. Cordasco[1], S. Marrone[2], F. Marulli[2], F. Cavallo[3], O. Gordeeva[4], Z. Callejas Carrión[5], A. Esposito[1]

[1]Dept of Psychology, Università degli Studi della Campania "Luigi Vanvitelli", and IIASS, Caserta and Vietri sul Mare, Italy;

[2]Dept of Mathematics and Physics (DMF), Università degli Studi della Campania "Luigi Vanvitelli"

[3]Dept of Industrial Eng, Florence University, BioRobotics Institute, SS Sant'Anna, and Dep of Excellence in Robotics & AI, Italy

[4]Acapela Group, Belgium

[5]Universidad de Granada, Spain

**Abstract.** The present study aims to explore user' acceptance and perceptions toward different quality levels of synthetical voices. To achieve this, four voices have been exploited considering two main factors: the quality of the voices (low vs high) and their gender (male and female). 186 volunteers were recruited and subsequently allocated into four groups of different ages respectively, adolescents, young adults, middle-aged and seniors. After having randomly listened to each voice, participants were asked to fill the Virtual Agent Voice Acceptance Questionnaire (VAVAQ). Outcomes show that the two higher quality voices of Antonio and Giulia were more appreciated than the low-quality voices of Edoardo and Clara by the whole sample in terms of pragmatic, hedonic and attractiveness qualities attributed to the voices. Concerning preferences towards differently aged voices, it clearly appeared that they varied according to participants age' ranges examined. Furthermore, in terms of suitability to perform different tasks, participants considered Antonio and Giulia equally adapt for healthcare and front office jobs. Antonio was also judged to be significantly more qualified to accomplish protection and security tasks, while Edoardo was classified as the absolute least skilled in conducting household chores.

## 1 Introduction

An overriding aim for the scientific research is to face challenges emerged from the increase of life expectancy. The World Health Organization (WHO) has clearly stated that it is a booming event compared to the past [1]. It is necessary to promote an autonomous and independent lifestyle for seniors inside their own homes, offering them sustainable and innovative solutions. The high number of requests, the costs and times foreseen by this type of personal assistance often cannot be satisfied by social and healthcare institutional services. Digital assistants could overcome all these limitations. To implement digital applications devoted to improving personal well-being, the characteristics of the end users have to be taken into account. Seniors are often considered as unaccustomed and inexperienced in the use of technology, and therefore it is essential that these devices are perceived as enthralling and easy to use [2]. Seniors' acceptance of mHealth solutions, described as based on "the use of mobile and wireless technologies to support the achievement of health goals"[3], could be undermined by their expectations of performance and effort, social influence, anxiety triggered by the use of technological devices and a proclivity to resist change [4]. It is not possible to better frame the issue of ageing by neglecting the recent health emergency due to the spread of coronavirus disease 2019 (COVID-19), which unfortunately continues to rage all over the world. This emergency has highlighted aspects of profound vulnerability that affect all aged groups of people and different contexts from various points of view. This unexpected change led to an abrupt awareness of the danger and the consequent need to adapt to heavy restrictions imposed by each government, inevitably upsetting people's daily routines. Social isolation and development of mental disorders represent two of the main consequences derived from interruptions in the daily routine [5,6]. In this context, a viable solution to reduce isolation and facilitate daily interactional exchanges could be the use of technological devices acting as assistants. Widely shared and spread is the burgeoning interest in chatbots, also known as conversational agents or virtual assistants. These digital tools may be suited for simulating conversations with humans. This is made possible thanks to machine learning and artificial intelligence methods used to replicate human-like behaviors and grant an activity-oriented dialogue structure. In terms of mental health research, several studies demonstrate the effectiveness of the therapeutic and preventive intervention provided by chatbots [7]. Noteworthy is Wisa, a chatbot able to elicit positive changes in its user's mood. The Touchkin eServices company has created this free mobile application to help people affected by depression and anxiety in managing their conditions [8]. Other studies have tested the effects of a chatbot focused on cognitive behavioral therapy. Outcomes of this survey report positive psychological interactions in a non-clinical population. [9]. Clearly, the use of voice is a fundamental tool to simplify the interaction with these devices. Some successful and commonly used application examples are represented by the hardware-based Amazon Echo running Alexa digital assistant software, the software-based Google Assistant running on Android devices. It appears that the use of synthetic voices helps to increase users' confidence when they are consumers. Due to a lack of data, it is not possible to say with equal confidence that the same occurs with users belonging to healthcare environments. Therefore, further investigation is needed. [10]. Studies exploring the role played by voice in agents' acceptance demonstrated that seniors prefer much more to interact with embodied speaking agents [11] and even with speaking not embodied visual interfaces [12], rather than voiceless agents embodied in a visual interface. These studies revealed that seniors strongly preferred to interact with speaking agents or even with agents' voices only rather than with mute agents and, moreover, voice affected seniors' positive assessment of the agents in terms of hedonic and pragmatic qualities. Additionally, female voices were preferred to male voices. This effect disappeared when both male and female agents were mute, revealing that gender preferences only occurred in the presence of agents' voice. Concerning young adults and adolescents, they did not seem to care whether the agent they were interacting with, was speaking or mute, female, or male. The quality of the synthetic voice has been however neglected; the current research aims to solve this issue. The investigations we are reporting concern the preferences of differently aged users (adolescents, young adults, middle-aged and senior participants) towards two different degrees of quality of synthetic voices (high and low quality) and synthetic voice' gender to draw up useful guidelines for the implementation of talking chatbots.

## 2  Materials and method

### 2.1  Participants

The present study involved 186 volunteers recruited in Campania (a Southern Italy region). Participants were divided in four groups equally balanced by gender. Group 1 was composed by 45 adolescents (24 males, mean age=15.13; SD=±0.79); Group 2 by 47 young adults (21 males, mean age=27.91; SD=±3.33); Group 3 by 45 middle-aged (22 males, mean age=48.87; SD=±4.11) and Group 4 by 49 seniors (23 males, mean age=75.22; SD=±8.02). The experiment was dedicated to investigating the degree of acceptance of four Italian synthetic voices represented by two female voices named Giulia and Clara and two male voices named Antonio and Edoardo. The voices differed not only in gender but also in two different levels of synthetic voice quality (low - Edoardo and Clara- vs high – Giulia and Antonio). In addition, participants were also asked to express their preferences regarding the age of the voices (preferred and attributed) and the proposed tasks for which they felt the voices might be more qualified. The whole sample joined the study on a voluntary basis and signed an informed consent formulated according to the current Italian and European laws about privacy and data protection. The research was approved by the ethical committee of the Università degli Studi della Campania "Luigi Vanvitelli," at the Department of Psychology, with the protocol number 25/2017.

**Stimuli**

Four synthetic voices lasting between 4 and 7 sec., were created for the experiment. Specifically, the low-quality voices (Edoardo and Clara) were created using the Natural Reader synthesizer (www.naturalreaders.com) and subsequently recorded with the free Audacity audio software (www.audacityteam.org). On the other hand, Acapela Group took care of the realization of the high-quality voices (Antonio and Giulia). This group is a European company based in Belgium, with over 30 years of leading experience in the production of high-quality synthetic voices (www.acapela-group.com). The voices were engineered as part of the H2020 funded Empathic project (www.empathic-project.eu) aimed at developing an empathic, expressive, and advanced virtual coach to assist seniors in their daily life. The assessment of the two different levels of voice quality was entrusted to Acapela's experts and to the BeCogSys laboratory team at the Università della Campania "Luigi Vanvitelli". Each voice reproduced the Italian sentence "Ciao sono Antonio/ Giulia/ Edoardo/ Clara. Se vuoi posso aiutarti nelle tue attività quotidiane" (Hi, my name is Antonio/ Giulia/ Edoardo/ Clara. If you allow me, I can assist you in your daily activities).

*Tools*

A digitalized version of the Virtual Agent Voice Acceptance Questionnaire (VAVAQ) was exploited to collect data and to assess participants' preferences toward the proposed synthetic voices. The VAVAQ derived from a previous version: the Virtual Agent Acceptance Questionnaire (VAAQ). This questionnaire has been developed inside the Empathic project to assess seniors' preferences toward virtual coach. Further details are reported in previous studies [13]. The digitalized version of VAVAQ was developed using Java scripts and allowed the automatic randomization of the questionnaire's items and sections presented to each participant. The questionnaire was structured as follows: an initial part of six items aimed at collecting socio-demographic information and eight following sections. Section 1 (composed of seven items) aimed at investigating participants' degree of experience and familiarity with technological devices such as smartphones, tablets, and laptops. Section 2 composed by a single item focused on participants' willingness to be involved in a long-lasting interaction with each proposed voice. Sections 3, 4, 5 and 6 (each consisting in ten items) assessed respectively: 1) the Pragmatic Qualities (PQ), i.e., the effectiveness, usefulness, practicality, clarity and controllability perceived by listening the voices; 2) the Hedonic-Identity Qualities (HQI), i.e., the originality, creativity and pleasantness attributed to the voices; 3) the Hedonic-Feeling Qualities (HQF) related to positive or negative arousals elicited by listening to the voices; 4) the Attractiveness (ATT) of the voices, i.e., the ability of the voice to engage listeners. Except for section 1, questionnaire's items required answers on 5-points Likert scale (1=strongly agree, 2=agree, 3=I do not know, 4=disagree, 5=strongly disagree). Questionnaire's items included either positive or negative statements, and scores from negative items were corrected in a reverse way. This implies that lower scores summon to more positive and high scores to less positive evaluations of the proposed voices. VAVAQ's section 7, constituted of three items, assessed which age participants were preferring (item 1) and attributing (item 3) to the listened voice along an age range from 1 (between 19-28 years old); 2 (between 29-38 years old); 3 (between 39-48 years old); 4 (between 49-58 years old); to 5 (59+ years old). The item 2 of section 7 explicitly asked participants whether the voice' age would affect their willingness to interact with them. The item contemplated only two possible answers: positive (yes) or a negative (no). In the current study, only participants' answers to item 1 of section 7 are reported. VAVAQ's section 8 consisted of four items exploring the occupations participants would entrust to the listened synthetic voices. Choices were among healthcare, housework, protection/security, and front office. This section required a 5-point Likert scale from 1=unsuitable, 2=hardly suitable, 3=I do not know, 4=quite suitable, to 5=very suitable and high scores reflected more positive voices' suitability for the proposed occupation.

*Procedures*

After being briefed on the study' aims, participants signed an informed consent. Then, they were asked to sit in front of a computer screen and fill the VAVAQ's section 1 for socio-demographic information. Subsequently, they randomly listened the four synthetic voices and after listening they were asked to fill the VAVAQ sections from 2 to 8. This procedure was repeated four times for each participant. No feedback or time limits were given to participants while reporting their answers.

## 3   Results

### 3.1   Data analysis

Separate ANOVA repeated measures analyses were carried out on the questionnaire's scores to assess participants' preferences toward voices' quality (high: Antonio and Giulia; low: Edoardo and Clara), age, and entrusted occupations. Participants' gender and their age group (adolescents= Group 1; young adults= Group 2; middle-aged= Group 3 and seniors= Groups 4) were considered as between subjects and VAVAQ scores as within subjects' factors. For sections 2 (willingness to interact), 3 (PQ), 4 (HQI), 5 (HQF) and 6 (ATT) due to the reverse correction of negative items, low scores summon to positive voices' assessments whereas high scores to negative ones. Additional ANOVA repeated measures analyses were performed on scores obtained from sections 7 and 8. Also in these cases, participants' gender and their age group were considered as between subjects and scores obtained from sections 7 and 8 (separately) as within subjects' factors. Scores from section 7 varied from 1 to 5 and high scores reflected participants' preferences toward more mature voices. Scores from section 8 also varied from 1 to 5 and high and low scores indicated respectively high and low suitability participants attributed to the four voices in accomplishing the proposed occupations. In all the analyses, the significance level was set at $\alpha < .05$ and differences among means were assessed through Bonferroni's post hoc tests. Details of abovementioned analyses are reported in the appendix. The main results are summarized below.

**Synthetic voices assessment**

The following section describes the results of the statistical analyzes reported in the appendix. These results reflect the preferences expressed by the four differently aged groups, towards the high- and low-quality male and female voices. To guide the interpretation of the scores shown graphically in figure 2, we invite you to observe that due to the reverse correction of negative items regarding the willingness to interact, the Pragmatic Qualities, the Hedonic Qualities- Identity (HQI) and Feeling (HQF) and the Attractiveness, low scores summon to positive voices' assessments whereas high scores to negative ones.

*Willingness to interact*

VAVAQ scores related to section 2 (willingness to interact) analyzed through the ANOVA analyses do not reveal significant differences for participants' gender. Significant differences emerged among age groups. In this context, seniors showed a greater willingness to interact with voices than adolescents, young adults, and middle-aged participants. In addition, willingness to interact scores significantly differed among the four proposed voices. Voice quality appeared to have been predominant in determining participants preferences rather than voice gender. High-quality voices of Antonio and Giulia respectively were significantly more appreciated than the low-quality voices corresponding to Edoardo and Clara. Deepening the statistical analysis, differences among the differently aged groups emerged for seniors which were significantly more willing to interact with high-quality male voice and low-quality both male and female voices than adolescents, young adults, and middle-aged participants. Regarding the high-quality female voice, it was found that seniors were more willing than adolescents and middle-aged participants whereas young adults expressed a stronger willingness to interact with the high-quality female voice than adolescents. Concerning significant differences in terms of willingness to interact with the four voices within each participants' group, young adults preferred the high-quality female voice to both male and female low-quality voices whereas they preferred high-quality male voice to the low- quality male voice. Moreover, middle-aged participants considered high-quality both female and male voices significantly more effective than the low-quality female and male voices. Figure 1 illustrates these results.

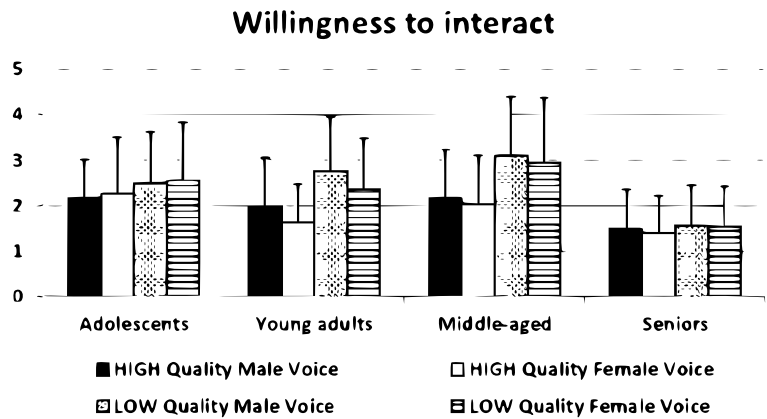

**Fig. 1.** Adolescents, young, middle-aged, and older adults' willingness to interact respectively with high quality male and female voices and low-quality male and female voices.

*Pragmatic Qualities*

The statistical analyses revealed that Pragmatic qualities (PQ) were not affected by participants' gender. Significant differences emerged among age groups. Specifically, seniors attributed to the proposed voices significantly higher PQ scores than the other three groups considered. PQ scores significantly differed among the four proposed voices. High- quality both male and female voices were significantly considered as more effective than the low-quality voices. Going into detail, differences among the differently aged groups emerged for seniors that considered high quality both female and male voices significantly more pragmatic than two specific groups (adolescents and middle-aged participants). However, when seniors were asked to evaluate low quality female and male voices, they attributed higher PQ scores to them than adolescents, young and middle-aged participants. Regarding significant differences in terms of PQ scores attributed to the four voices, within each participants' group, analyses revealed that both young adults and middle-aged participants attributed to high quality male and female voices better PQ scores than those attributed to low-quality female and male voices. Figure 2 illustrates these results.

*Hedonic Qualities- Identity (HQI)*

Hedonic Qualities- Identity (HQI) were not affected by participants' gender. Significant differences emerged among age groups. Essentially, seniors attributed to the four proposed voices significantly higher HQI scores than adolescents, young and middle-aged participants. HQI scores significantly differed among the four proposed voices. High-quality both male and female voices were significantly considered as more original, creative, and pleasant than the low-quality male and female voices. Significant differences among the differently aged groups emerged for seniors that appreciated the high-quality male voice and both female and male low-quality voices significantly more than adolescents, young and middle-aged participants and they better assessed the high-quality female voice than adolescents and middle-aged participants. Regarding significant differences in terms of HQI scores, within each participants' group, analyses revealed that both young adults and middle-aged participants attributed to high quality male and female voices better HQI scores than those attributed to low-quality female and male voices. Figure 2 illustrates these results.

*Hedonic Qualities- Feeling (HQF)*

Hedonic Qualities- Feeling (HQF) were not affected by participants' gender. Significant differences emerged among age groups. Seniors attributed to the four proposed voices significantly higher HQF scores than adolescents, young and middle-aged participants. HQF scores significantly differed among the four proposed voices. High-quality both male and female voices elicited significantly more positive arousals than the low-quality male and female voices. Regarding significant differences in terms of HQF scores attributed to the four voices, within each participants' group, analyses revealed that young adults attributed to high quality female voice better HQF scores than those attributed to low-quality female and male voices; they also considered the high-quality male voice more engaging than the low-quality male voice. In addition, middle-aged adults attributed to high quality male and female voices better HQI scores than those attributed to low-quality female and male voices. Figure 2 summarizes these results.

*Attractiveness*

Attractiveness of voices was not affected by participants' gender. Significant differences emerged among age groups. Seniors attributed significantly more attractiveness to the four proposed voices rather than adolescents, young and middle-aged participants. High-quality both male and female voices were considered significantly more attractive than the low-quality male and female voices. Regarding significant differences in terms of ATT scores attributed to the four voices, within each participants' group, analyses revealed that young adults attributed to high quality female voice better ATT scores than those attributed to low- quality female and male voices; they also considered the high- quality male voice more engaging than the low-quality male voice. In addition, middle-aged adults attributed to high quality male and female voices better ATT scores than those attributed to low-quality female and male voices. Figure 2 illustrates these results.

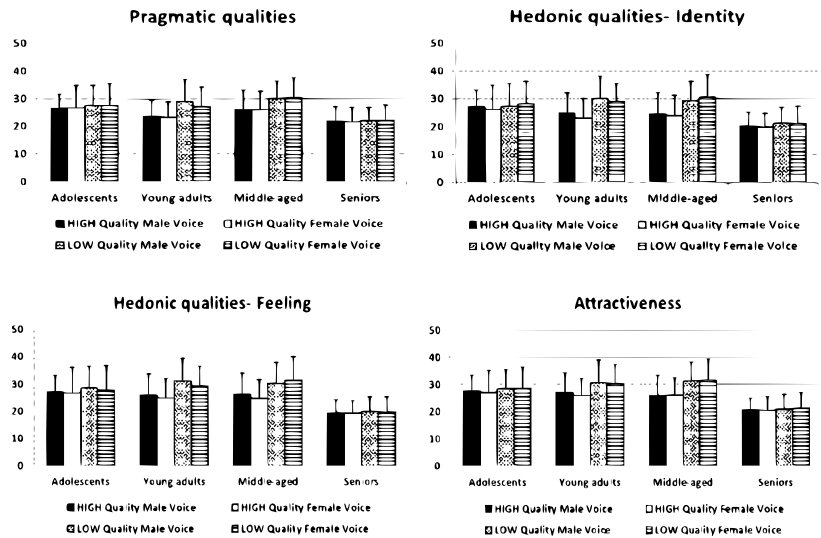

**Fig. 2.** PQ, HQI, HQF and ATT scores attributed by adolescents, young, middle-aged, and older adults respectively to high quality male and female voices and low-quality male and female voices.

*Preferred age range results*

This descriptive section summarizes the percentage values of the synthetic voices age preferences computed for adolescents, young adults, middle-aged and older participants as exemplified in Table 1. Details of the statistical analyses are reported in the appendix.

The preferred age was distributed according to the following percentages among the differently aged groups. Data revealed that 53.33% of adolescents (aged between 14-16 years) seemed to prefer the age range between 19-28 years while 36.11% of them selected the age range between 29-38 years suggesting that not having the possibility to select among the different options of choice (1= 19-28 years; 2= 29-38 years; 3= 39-48 years; 4= 49-58 years and 5= 59+ years) their own reference age group, they selected the age ranges closest to theirs. Young adults (aged between 22-35 years) mostly showed a stronger preference for the age range between 29-38 years (55.85% of young adults) and then for the two age ranges positioned immediately before and after it, namely those between 19-28 and 39-48 years (23.4% and 18.62% of them, respectively). Regarding middle-aged participants (aged between 40-55 years), they disclosed their preferences towards synthetic voices that matched their own age range, mainly by choosing the age range between 39-48 years (46.67% of the participants' group). Interestingly, 36.11% of them expressed a clear preference for the age range between 29-38 years by preferring younger voices. Finally, the participants' group of seniors (65+ years old) was the only one who revealed a particularly heterogeneous distribution of preferences. In this context, 26.02% of seniors preferred voices of their same age range (59+), 27.55% of them selected the age range between 29-38 years, 19.9% preferred the age range between 39-48 and 19.9% indicated to prefer the age range between 49-58 years.

Table 1. Preferred age range percentage values attributed to synthetic voices by the four differently aged groups of participants. Scores varied from 1 to 5 and reflected age ranges (1=19-28 years old; 2=29-38 years old; 3=39-48 years old; 4=49-58 years old; 5=+59 years old).

| % Age preference | 19-28 years | 29-38 years | 39-48 years | 49-58 years | 59+ |
|---|---|---|---|---|---|
| **Adolescents** | 53.33% | 36.11% | 8.33% | 2.22% | 0% |
| **Young adults** | 23.4% | 55.85% | 18.62% | 1.6% | 0.53% |
| **Middle-aged** | 9.44% | 36.11% | 46.67% | 7.78% | 0% |
| **Seniors** | 6.63% | 27.55% | 19.9% | 19.9% | 26.02% |

*Entrusted occupations to the synthetic voices*

In the following are reported the suitability scores attributed to the voices to healthcare, housework, protection and security task, and front-office occupations by the four differently aged groups of participants. However, it must be clear that participants were able to entrust different suitability scores to the voices for the different proposed occupations only by listening to them. To better interpret the scores shown graphically in figure 3, we invite you to observe that high and low scores indicated respectively high and low suitability participants attributed to the four voices in accomplishing the proposed occupations.

*Healthcare*

The statistical analyses revealed that healthcare occupations were not affected by participants' gender. Significant differences emerged among age groups. Seniors considered voices as significantly more suitable than adolescents, young and middle-aged participants to perform healthcare occupations. Suitability scores for healthcare occupations significantly differed among the four proposed voices. High- quality both male and female voices were considered significantly as more appropriate than the low-quality male and female voices for healthcare occupations. Moreover, adolescents better rated low quality both female and male voices than middle-aged participants in performing healthcare occupations. Regarding significant differences in terms of healthcare scores attributed to the four voices, within each participants' group, analyses revealed that young adults attributed to high quality female voice better scores than those attributed to low-quality both female and male voices. In addition, middle-aged adults better assessed high quality both female and male voices in performing healthcare task rather than low-quality female and male voices. Figure 3 illustrates these results.

*Housework*

Housework tasks were not affected by participants' gender. Significant differences emerged among age groups. Seniors considered voices as significantly more appropriate than adolescents, young and middle-aged participants in performing housework tasks. Additionally, young adults considered the proposed voices more suitable than middle-aged participants to housework occupations. Significant gender differences among the differently aged groups emerged. Specifically, for two groups: male seniors and female middle-aged participants. Male seniors attributed to the voices higher suitability scores than male adolescents, male young adults, and male middle-aged participants in performing housework occupations while female middle-aged participants attributed to the voices lower suitability scores than female young and female older adults. Interestingly, statistical analyses revealed a significant gender difference within seniors' age group. Male seniors considered the voices as more suited than female older adults in performing housework tasks. The four proposed voices differed significantly in their suitability scores to perform housework. Low quality male voice was rated significantly less suited than low quality female voice and high quality both male and female voices for housework occupations. Significant differences among the differently aged groups emerged for seniors that attributed to high quality male voice and low-quality male voice higher suitability scores in performing housework than adolescents, young and middle- aged participants. Furthermore, seniors better assessed high quality female voice and low-quality female voice as more suitable than adolescents and middle-aged participants. Additionally, young adults rated low quality male voice more appropriate than middle-aged adults and young adults attributed higher suitability scores low quality female voice than adolescents and middle-aged participants. Regarding significant differences in terms of housework scores attributed to the four voices, within each participants' group, analyses revealed that middle-aged participants

11attributed to high quality female voice better scores than those attributed to high quality male voice and low-quality both female and male voices. In addition, middle-aged adults better assessed high quality male voice in performing housework task rather than low-quality male voice. Figure 3 shows these results.

*Protection and security tasks*

Protection and security tasks were not affected by participants' gender. Significant differences emerged among age groups. Seniors considered the voices significantly more suitable than adolescents, middle-aged participants in performing protection and security tasks. Suitability scores for protection and security tasks differed significantly among the four proposed voices. High quality male voice was considered significantly more qualified than low quality both male and female voices for protection and security tasks. Significant differences among the differently aged groups emerged for seniors that considered high quality male voice significantly more qualified than adolescents in performing these specific tasks. Moreover, seniors also considered low quality male voice as more suited than middle-aged participants and they rated low quality female voice as more suitable than adolescents, young and middle-aged participants. Additionally, middle-aged participants considered low quality female voice significantly less qualified for protection and security tasks than adolescents and young adults. Regarding significant differences in terms of protection and security tasks scores attributed to the four voices, within each participants' group, analyses revealed that young adults attributed to high quality male voice better scores than those attributed to low-quality female and high-quality female voices. In addition, middle- aged adults better assessed high quality both female and male voices in performing protection and security tasks rather than low-quality female and male voices. Figure 3 illustrates these results.

*Front office*

Front office tasks were not affected by participants' gender effect and there were not significant differences among age groups. Suitability to front office tasks was rated significantly different among the four proposed voices. High quality both female and male voices were considered significantly more suitable than low quality both female and male voices for front office tasks. Significant differences among the differently aged groups emerged for middle-aged participants that rated high quality female voice as more qualified in performing front office tasks than seniors. Regarding significant differences in terms of front office scores attributed to the four voices, within each participants' group, analyses revealed that young adults attributed to high quality both female and male voices better scores than those attributed to low-quality both female and male voices. In addition, middle-aged adults better assessed high quality both female and male voices in performing front office tasks rather than low-quality both female and male voices. Figure 3 shows these results.

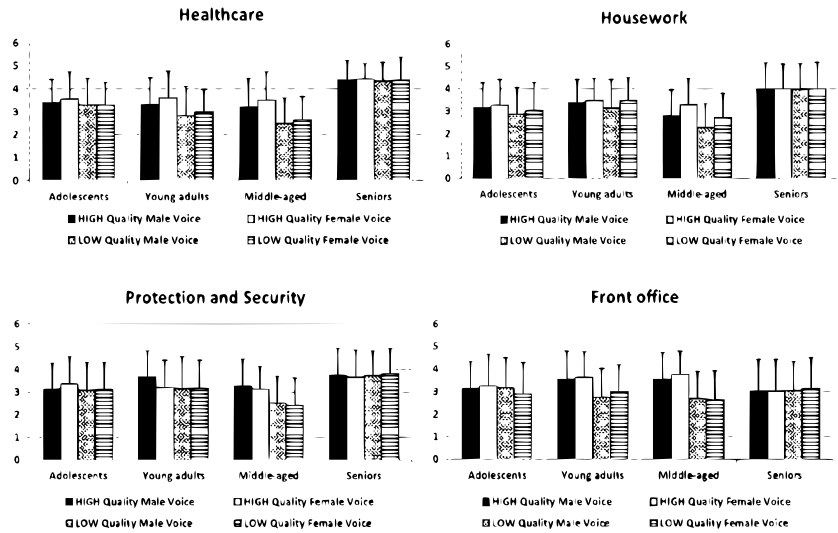

**Fig. 3.** Suitability scores attributed to the voices in performing healthcare, housework, protection and security and front office tasks.

## 4      Discussion and conclusions

The present study explored which synthetic voices' features could influence the way people interact with them. Outcomes revealed that seniors show a greater willingness to be involved in a long-lasting interaction with the four proposed voices regardless of the quality of their synthesis. Seniors' appreciation of voices was also envisaged from the scores they attributed to the voices' pragmatic, hedonic, and attractive qualities. The fact that seniors did not care too much of the voices' quality may reflect their low experience with technology and a low accuracy in detecting the different qualities of the proposed voices. On the other hand, these results certainly show a greater seniors' propensity to potentially receive assistance from automatic devices explicating their human abilities only by voices. This result was not observed for the other differently age groups involved in this experiment. Concerning adolescents, young adults, and middle-aged participants, they seem to be more selective in evaluating the four synthetic voices, suggesting that the new generations are more skillful and accustomed to detecting subtle differences in the voice quality of the proposed devices and call for improved sound's resolution and definition. It was observed, that for all the investigated (PQ, HQI, HQF, and ATT) qualities, high quality voices (respectively, Antonio and Giulia) were significantly more appreciated than low quality voices (Edoardo and Clara). The two high quality voices were judged significantly more practical, controllable, pleasant, original, able to arouse positive feelings and engage users in an effective long-lasting interaction compared to the low-quality synthetic speech of Edoardo and Clara. The lesson learned by these data is that people are more sensible to the voice' quality of their assistive devices rather than to their gender. Antonio and Giulia's voices have both received greater acclaim regardless to their gender. As regard the statistical analyses (widely described in appendix section) related to users' preferred age of their assistive devices, it appeared clear that preferences vary with the users' age. Adolescents and young adults feel more pleased to be assisted by peer aged voices, middle-aged participants prefer to be assisted by more mature voices and seniors by even more mature voices than the preferences expressed by middle-aged participants. However, middle-aged and seniors preferred age of their assistive devices is far from their own age. Interestingly, participants' age preferences toward Giulia and Clara's voices, were collocated in the same average age ranges. Focusing on the occupations participants would have entrusted to the proposed synthetic voices, the evaluations took on a more specific connotation. Seniors considered voices significantly more qualified than adolescents, young adults, and middle-aged participants in performing healthcare, housework, and protection/security tasks. Additionally, young adults attributed to the voices higher ability to successfully accomplish housework than middle-aged participants. Antonio and Giulia's high quality synthetic voices were equally perceived as more suited than Edoardo and Clara for healthcare and front office jobs. Besides, Antonio was judged significantly more qualified for protection and security tasks than Edoardo and Clara's voices and surprisingly, Edoardo's voice ranked as the last one in terms of suitability to housework. Future work must include clinical populations. It is hoped that these findings can be exploited to guide the design and the implementation of chatbots' voices devoted to providing support and intervention in healthcare settings.

**ACKNOWLEDGMENT**

The research leading to these results has received funding from the European Union Horizon 2020 research and innovation programme under grant agreement N. 769872 (EMPATHIC) and N. 823907 (MENHIR), from the project SIROBOTICS that received funding from Ministero dell'Istruzione, dell 'Università, e della Ricerca (MIUR), PNR 2015-2020, Decreto Direttoriale 1735 July 13 2017, and from the project ANDROIDS that received funding from Università della Campania "Luigi Vanvitelli" inside the programme V:ALERE 2019, funded with D.R. 906 del 4/10/2019, prot. n. 157264, October 17, 2019

## References

[1] Ageing and Health, https://www.who.int/news-room/fact-sheets/detail/ageing-and-health, accessed on 30 November 2020.

[2] Mannheim, I., Schwartz, E., Xi, W., Buttigieg, S. C., McDonnell-Naughton, M., Wouters, E. J., & Van Zaalen, Y.: Inclusion of older adults in the research and design of digital technology. International journal of environmental research and public health, 16(19), 3718 (2019).

[3] World Health Organization, "mHealth: new horizons for health through mobile technologies," mHealth: new horizons for health through mobile technologies, (2011).

[4] Hoque, R., & Sorwar, G.: Understanding factors influencing the adoption of mHealth by the elderly: An extension of the UTAUT model. International journal of medical informatics, 101, 75-84 (2017).

[5] Liu, J. J., Bao, Y., Huang, X., Shi, J., & Lu, L.: Mental health considerations for children quarantined because of COVID-19. The Lancet Child & Adolescent Health, 4(5), 347-349 (2020).

[6] Lyall, L. M., Wyse, C. A., Graham, N., Ferguson, A., Lyall, D. M., Cullen, B., ... & Smith, D. J.: Association of disrupted circadian rhythmicity with mood disorders, subjective well-being, and cognitive function: a cross-sectional study of 91 105 participants from the UK Biobank. The Lancet Psychiatry, 5(6), 507-514 (2018).

[7] Suganuma, S., Sakamoto, D., & Shimoyama, H.: An embodied conversational agent for unguided internet-based cognitive behavior therapy in preventative mental health: feasibility and acceptability pilot trial. JMIR mental health, 5(3), e10454 (2018).

[8] Inkster, B., Sarda, S., & Subramanian, V.: An empathy-driven, conversational artificial intelligence agent (Wysa) for digital mental well-being: real-world data evaluation mixed-methods study. JMIR mHealth and uHealth, 6(11), e12106 (2018).

[9] Ly, K. H., Ly, A. M., & Andersson, G.: A fully automated conversational agent for promoting mental well-being: A pilot RCT using mixed methods. Internet interventions, 10, 39-46 (2017).

[10] Qiu, L., & Benbasat, I.: Online consumer trust and live help interfaces: The effects of text-to-speech voice and three-dimensional avatars. International journal of human-computer interaction, 19(1), 75-94 (2005).

[11] Esposito, A., Amorese, T., Cuciniello, M., Riviello, M. T., Esposito, A. M., Troncone, A., ... & Cordasco, G.: Elder user's attitude toward assistive virtual agents: the role of voice and gender. Journal of Ambient Intelligence and Humanized Computing, 12(4), 4429-4436 (2021).

[12] Esposito, A., Amorese, T., Cuciniello, M., Riviello, M. T., Esposito, A. M., Troncone, A., & Cordasco, G.: The Dependability of Voice on Elders' Acceptance of Humanoid Agents. In Interspeech, pp. 31-35 (2019, September).

[13] Esposito, A., Amorese, T., Cuciniello, M., Esposito, A. M., Troncone, A., Torres, M. I., ... & Cordasco, G.: Seniors' acceptance of virtual humanoid agents. In Italian forum of ambient assisted living, pp. 429-443. Springer, Cham (2018, July).

## Appendix: detailed description of the statistical analyses performed for VAVAQ data

Separate ANOVA repeated measures analyses were performed on the questionnaire's scores to evaluate participants' preferences toward voices' quality (high: Antonio and Giulia; low: Edoardo and Clara), ages, and entrusted occupations. Participants' gender and their age group (adolescents= Group 1; young adults= Group 2; middle-aged= Group 3 and seniors= Groups 4) were considered as between subjects and VAVAQ scores as within subjects' factors. For sections 2 (willingness to interact), 3 (PQ), 4 (HQI), 5 (HQF) and 6 (ATT) due to the reverse correction of negative items, low scores summon to more positive whereas high scores to less positive voices' assessments. Separate ANOVA repeated measures analyses were conducted on scores obtained from sections 7 and 8. Also in these cases, participants' gender and their age group were considered as between subjects and scores obtained from sections 7 and 8 (separately) as within subjects' factors. Scores from section 7 varied from 1 to 5 and high scores reflected participants' preferences toward more mature synthetic voices. Scores from section 8 also varied from 1 to 5 and high and low scores indicated respectively high and low suitability participants attributed to the four voices in accomplishing the proposed occupations. In all the analyses, the significance level was set at α < .05 and differences among means were assessed through Bonferroni's post hoc tests. The analyses' results are reported below.

**Synthetic voices assessment**

In the following section are reported the preferences expressed by the four differently aged groups, toward the synthetic voices. Henceforth we will use only the name of the voices. Specifically, high quality both female and male voices were belonging respectively to Giulia and Antonio while low quality female and male voices were identified by Clara and Edoardo' voices, respectively. However, it must be clear that participants were able to evaluate the synthetic voices by answering questionnaire sections items only by listening to them. Please note, due to the reverse correction of negative items, low scores summon to positive voices' assessments whereas high scores to negative ones.

*Willingness to interact*

No significant effects of participants' gender ($F_{(1,178)} = 2.167$, $p=.143$) was observed for the willingness to interact with the synthetic voices. Significant differences emerged among age groups ($F_{(3,178)} = 20.370$, $p<<.01$). Bonferroni's post hoc tests revealed that seniors' (mean=1.486) willingness to interact with the proposed voices was significantly stronger than adolescents (mean= 2.366, $p<<.01$), young adults (mean=2.194, $p<<.01$) and middle-aged participants (mean=2.563, $p<<.01$). Willingness to interact scores significantly differed among the four proposed voices ($F_{(3,534)} = 20.665$ $p<<.01$). Bonferroni's post hoc tests revealed that this difference was due to Giulia (mean=1.831, $p<<.01$) and Antonio's (mean=1.960, $p<<.01$) voices which were significantly more appreciated than Edoardo (mean=2.476) and Clara's (mean=2.341) voices.

A significant interaction emerged between age groups and participants' willingness to interact with the synthetic voices ($F_{(9,534)} = 3.356$, $p=.001$). Bonferroni's post hoc tests were performed for each single factor (age groups and willingness to interact scores). Tests revealed that:

a) Concerning the age groups, seniors (mean=1.479) were significantly more willing to interact than adolescents (mean=2.173, $p=.003$), young adults (mean=2.014, $p=.042$) and middle-aged (mean=2.173, $p=.003$) participants with Antonio's voice. Similarly, seniors (mean=1.395) were significantly more willing to interact than adolescents (mean=2.259, $p<<.01$) and middle-aged (mean=2.036, $p=.010$) participants with Giulia's voice. Young adults (mean=1.636) expressed a greater willingness than adolescents (mean=2.259, $p=.016$) with Giulia's voice. Edoardo's voice was significantly more appreciated by seniors (mean=1.554) than adolescents (mean=2.485, $p=.001$), young adults (mean=2.756, $p<<.01$) and middle-aged participants (mean=3.109, $p<<.01$). Finally, Clara's voice was significantly more appreciated by seniors (mean=1.515) than adolescents (mean=2.548, $p<<.01$), young adults (mean=2.368, $p=.004$) and middle-aged (mean=2.933, $p<<.01$) participants.

b) Concerning the willingness to interact scores Bonferroni's post hoc tests revealed that young adults' preferences toward Giulia's voice (mean=1.636) were significantly more influential than their preferences towards Edoardo (mean=2.756, $p\ll.01$) and Clara's (mean=2.368, p=.003) voices. In addition, young adults' preferences toward Antonio's (mean=2.014) voice were significantly more effective than their preferences toward Edoardo's (mean=2.756, $p\ll.01$) voice. Concerning middle-aged participants, their preferences toward Giulia (mean=2.036) and Antonio's (mean=2.173) voices were significantly more effective that their preferences toward Edoardo (mean=3.109, $p\ll.01$) and Clara's (mean=2.933, $p\ll.01$) voices.

*Pragmatic Qualities*

Pragmatic qualities (PQ) were not affected by participants' gender ($F(1,178)$ = .147, p=.702). Significant differences emerged among age groups ($F(3,178)$ = 16.502, $p\ll.01$). Bonferroni's post hoc tests revealed that seniors (mean=21.943) attributed to the proposed synthetic voices significantly higher PQ scores than adolescents (mean=27.052, $p\ll.01$), young adults (mean=25.853, $p\ll.01$) and middle-aged participants (mean=28.197, $p\ll.01$). PQ scores significantly differed among the four proposed voices ($F(3,534)$ = 13.583, $p\ll.01$). Bonferroni's post hoc tests revealed that PQ scores attributed to Giulia (mean=24.458) and Antonio's (mean= 24.577) voices were significantly more positive that PQ scores attributed to Edoardo (mean=27.189, $p\ll.01$) and Clara's (mean=26.821, $p\ll.01$) voices.

A significant interaction emerged between age groups and PQ scores ($F(9,534)$ = 3.017, p=.002). Bonferroni's post hoc tests were performed for each single factor (age groups and PQ scores). Tests revealed that:

a) Concerning age groups, seniors (mean=21.930) considered Antonio's voice significantly more pragmatic than adolescents (mean=26.455, p=.001) and middle-aged (mean=26.097, p=.004) participants. Seniors (mean=21.747) also considered Giulia's voice significantly more pragmatic than adolescents (mean=26.661, p=.002) and middle-aged (mean=26.106, p=.008) participants. Seniors (mean=22.044) considered Edoardo's voice significantly more pragmatic than adolescents (mean=27.545, p=.001), young (mean=29.110, $p\ll.01$) and middle-aged participants (mean=30.056, $p\ll.01$). Finally, seniors (mean=22.052) considered Clara's voice significantly more pragmatic than adolescents (mean=27.548, p=.001), young (mean=27.157, p=.003) and middle-aged (mean=30.528, $p\ll.01$) participants.

b) Regarding PQ scores, PQ scores attributed by young adults to Antonio (mean=23.827) and Giulia's voices (mean=23.320) were significantly higher than PQ scores attributed to Edoardo (mean=29.110, $p\ll.01$) and Clara's voices (mean=27.157, p<.05). PQ scores attributed by middle-aged adults to Giulia (mean=26.106) and Antonio's voices (mean=26.097) were significantly higher than PQ scores attributed to Edoardo (mean=30.056, $p\ll.01$) and Clara's voices (mean=30.528, $p\ll.01$).

*Hedonic Qualities- Identity (HQI)*

Hedonic Qualities- Identity (HQI) were not affected by participants' gender ($F(1,178)$ = .023, p=.879). Significant differences emerged among age groups ($F(3,178)$ = 25.426, $p\ll.01$). Bonferroni's post hoc tests revealed that seniors (mean=20.567) attributed to the four proposed voices significantly higher HQI scores than adolescents (mean=27.300, p<<.01), young (mean=27.101, p<<.01) and middle-aged participants (mean=27.135, p<<.01). HQI scores significantly differed among the four proposed voices ($F_{(3,534)}$ =20.223, p<<.01). Bonferroni's post hoc tests revealed that Antonio (mean=24.330) and Giulia's (mean=23.237) voices were significantly more appreciated than Edoardo (mean=27.195, p<<.01) and Clara's (mean=27.341, p<<.01) voices.

A significant interaction emerged between age groups and HQI scores ($F_{(9,534)}$ = 2.953, p=.002). Bonferroni's post hoc tests were performed for each single factor (age groups and HQI scores). Tests revealed that:

a) Concerning age groups, seniors (mean=20.157) considered Antonio's voice significantly more original, creative, and pleasant than adolescents (mean=27.274, p<<.01), young (mean=25.308, p=.001) and middle-aged (mean=24.581, p=.008) participants. Furthermore, seniors (mean=19.623) considered Giulia's voice as significantly more pleasant than adolescents (mean=26.170, p<<.01) and middle-aged participants (mean=23.921, p=.025). Edoardo's voice was more appreciated by seniors (mean=21.439) than adolescents (mean=27.443, p=.001), young (mean=30.478, p<<.01) and middle-aged participants (mean=29.419, p<<.01). Hedonic qualities-identity assessment revealed that seniors (mean=21.048) evaluated Clara's voice as significantly more original, creative, and pleasant than adolescents (mean=28.313, p<<.01), young (mean=29.385, p<<.01) and middle-aged participants (mean=30.618, p<<.01).

b) Concerning the HQI scores, young adults attributed to Giulia (mean=23.235) and Antonio's (mean=25.308) voices significantly higher HQI scores than those attributed to Edoardo (mean=30.478, p<<.01) and Clara's voices (mean=29.385, p<<.01). Similarly, HQI scores attributed by middle-aged participants to Giulia (mean=23.921) and Antonio's (mean=24.581) voices were significantly higher than those they attributed to Edoardo (mean=29.419, p<<.01) and Clara's (mean=30.618, p<<.01) voices.

*Hedonic Qualities- Feeling (HQF)*

Hedonic Qualities- Feeling (HQF) were not affected by participants' gender ($F_{(1,178)}$ = .346, p=.557) Significant differences emerged among age groups ($F_{(3,178)}$ = 34.836, p<<.01). Bonferroni's post hoc tests revealed that seniors (mean=19.414) HQF were significantly higher than HQF scores attributed to the four proposed voices by adolescents (mean=27.429, p<<.01), young (mean=27.881, p<<.01) and middle-aged participants (mean=28.103, p<<.01). HQF scores significantly differed among the four proposed voices ($F_{(3,534)}$ =14.838, p<<.01). Bonferroni's post hoc tests revealed that Antonio (mean=24.612) and Giulia's voices (mean=23.827) elicited significantly more positive arousals than Edoardo (mean=27.452, p<<.01) and Clara's (mean=26.935, p<<.01) voices.

A significant interaction emerged between age groups and HQF scores ($F_{(9,534)}$ = 2.927, p= .002). Bonferroni's post hoc tests were performed for each single factor (age groups and HQF scores). Tests revealed that:

a) Concerning age groups, seniors (mean=19.143) considered Antonio's voice significantly more engaging than adolescents (mean=27.054, p<<.01), young (mean=26.137, p<<.01) and middle-aged (mean=26.114, p<<.01) participants. Seniors (mean=19.246) also considered Giulia's voice significantly more engaging than adolescents (mean=26.542, p<<.01), young (mean=24.858, p=.001) and middle-aged participants (mean=24.664, p=.002). Similarly, seniors (mean=19.765) considered Edoardo's voice significantly more able to elicit positive feelings than adolescents (mean=28.551, p<<.01), young (mean=31.238, p<<.01) and middle-aged participants (mean=30.255, p<<.01). Finally, seniors (mean=19.502) evaluated Clara's voice significantly better than adolescents (mean=27.568, p<<.01), young (mean=29.292, p<<.01) and middle-aged participants (mean=31.377, p<<.01).

b)Concerning HQF scores, young adults attributed to Giulia's (mean=24.858) voice significantly higher HQF scores than those they attributed to Edoardo (mean=31.238, p<<.01) and Clara's voices (mean=29.292, p=.010), and attributed to Antonio's voice (mean=26.137) significantly higher HQF scores than those they attributed to Edoardo's voice (mean=31.238, p<<.01). In addition, middle-aged adults attributed to Giulia (mean=24.664) and Antonio's voices (mean=26.114) significantly higher HQF scores than those they attributed to Edoardo (mean=30.255, p<<.01) and Clara's (mean=31.377, p<<.01) voices.

*Attractiveness*

Attractiveness of synthetic voices was not affected by participants' gender ($F(1,178) = .005$, p=.944). Significant differences emerged among age groups ($F(3,178) = 31.066$, p<<.01). Bonferroni's post hoc tests revealed that seniors (mean=20.775) attributed significantly more attractiveness to the four proposed voices rather than adolescents (mean=27.851, p<<.01), young (mean=28.568, p<<.01) and middle-aged participants (mean=28.710, p<<.01). ATT scores significantly differed among the four proposed voices ($F(3,534) =15.737$, p<<.01). Bonferroni's post hoc tests revealed that Antonio (mean=25.337) and Giulia's (mean=24.781) voices were considered significantly more attractive than Edoardo (mean=27.821, p<<.01) and Clara's (mean=27.966, p<<.01) voices.

A significant interaction emerged between age groups and ATT scores ($F(9,534) = 2.910$, p=.002). Bonferroni's post hoc tests were performed for each single factor (age groups and ATT scores). Tests revealed that:

a)Concerning age groups, seniors (mean=20.623) considered significantly more attractive than adolescents (mean=27.643, p<<.01), young (mean=27.313, p<<.01) and middle-aged (mean=25.768, p=.001) participants Antonio's voice. Seniors (mean=20.406) considered significantly more attractive than adolescents (mean=26.866, p<<.01), young (mean=25.826, p=.001) and middle-aged (mean=26.027, p<<.01) participants Giulia's voice. Seniors (mean=20.863) considered significantly more attractive than adolescents (mean=28.408, p<<.01), young (mean=30.701, p<<.01) and middle-aged participants (mean=31.310, p<<.01) Edoardo's voice. Finally, seniors (mean=21.207) considered significantly more attractive than adolescents (mean=28.488, p<<.01), young (mean=30.432, p<<.01) and middle-aged participants (mean=31.737, p<<.01) Clara's voice.

b)Concerning ATT scores, Giulia's voice (mean=25.826) received by young adults significantly higher ATT scores than Edoardo (mean=30.701, p<<.01) and Clara's (mean=30.432, p=.003) voices, and Antonio's voice (mean=27.313) received significantly higher ATT scores than Edoardo's (mean=30.701, p=.013) voice. Giulia (mean=26.027) and Antonio's (mean=25.768) voices received by middle-aged participants significantly higher ATT scores than Edoardo (mean=31.310, p<<.01) and Clara's (mean=31.737, p<<.01) voices.

*Preferred age range results*

Data relating to the voices' age preferences are collected through section 7 of the digitized version of the VAVAQ. This section consists of three items. The third item specifically asks participants to express their voice preferred age range among the following: 1= 19-28 years; 2= 29-38 years; 3= 39-48 years; 4= 49-58 years and 5= 59+ years. Therefore, scores from 1 to 5 are not real scores but indicate different age ranges.

Preferred age range was not affected by participants' gender (F (1,178) = .000, p= .987) while it was significantly affected by age groups (F (3,178) = 44.334, p<<.01). Bonferroni post hoc test revealed that middle-aged participants (mean=2.525) differed significantly from adolescents (mean=1.601, p<<.01) and young adults (mean=1.989, p=.007) by preferring more aged voices. Similarly, older adults (mean=3.331) significantly differed from adolescents (mean=1.601, p<<.01), young (mean=1.989, p<<.01) and middle-aged participants (mean=2.525, p<<.01) preferring even more mature voices.

A significant interaction emerged between participants' gender and age groups (F (3,178) = 3.707, p=.013). Bonferroni's post hoc tests were performed for each single factor (age groups and participants' gender). Tests showed that:

a) Concerning age groups age preferences expressed by male seniors (mean=3.652) differed significantly from age preferences expressed by male adolescents (mean=1.500, p<<.01), male young (mean=1.881, p<<.01) and male middle-aged adults (mean=2.409, p<<.01). Age preferences expressed by male adolescents (mean=1.500) differed significantly from age preferences expressed by male middle-aged (mean=2.409, p=.001) adults. Male adolescents preferred voices closer to their age range while male middle-aged participants preferred more aged voices even though younger than their own age. Female older (mean=3.010) adults significantly differed from female adolescents (mean=1.702, p<<.01) and female young adults (mean=2.096, p<<.01) in their age preferences. Female adolescents' (mean=1.702) age preferences differed significantly from female middle-aged participants' (mean=2.641, p=.001) age preferences.

b) Concerning participants' gender, age preferences expressed by male older adults (mean=3.652) significantly differed from age preferences expressed by female older adults (mean=3.010, p=.004).

*Entrusted occupations to the synthetic voices*

In the following are reported the suitability scores attributed to the synthetic voices (only by listening to them) to healthcare, housework, protection and security task, and front-office occupations by the four differently aged groups of participants. Please note, scores from this section varied from 1 to 5 and high and low scores indicated respectively high and low suitability that participants attributed to the four voices in accomplishing the proposed occupations.

*Healthcare*

No participants' gender effect (F (1,178) = .911, p=.341) was found. Significant differences among age groups (F (3,178) = 30.994, p<<.01) emerged. Bonferroni's post hoc tests revealed that seniors (mean=4.391) considered the voices as significantly more suitable than adolescents (mean=3.387, p<<.01), young (mean=3.187, p<<.01) and middle-aged participants (mean=2.964, p<<.01) to perform healthcare occupations. Suitability scores for healthcare occupations significantly differed among the four proposed voices (F (3,534) = 14.075, p<<.01). Bonferroni's post hoc tests revealed that Antonio's voice (mean=3.575) was considered as more appropriate than Edoardo (mean=3.240, p=.002) and Clara's voices (mean=3.330, p=.043), and Giulia's (mean=3.784) voice as more suited than Edoardo (mean=3.240, p<<.01) and Clara's (mean=3.330, p<<.01) voices for healthcare occupations.

A significant interaction emerged between age groups and the synthetic voices' suitability to provide healthcare assistance (F (9,534) = 2.416, p=.011), Bonferroni's post hoc tests were performed for each single factor (age groups and healthcare scores). Tests revealed that:

a) Concerning age groups, Antonio received higher suitability scores in performing healthcare occupation by seniors (mean=4.393) rather than adolescents (mean=3.396, p<<.01), young (mean=3.307, p<<.01) and middle-aged adults (mean=3.205, p<<.01). Giulia received higher suitability scores by seniors (mean=4.436) rather than adolescents (mean=3.571, p=.001), young (mean=3.608, p=.001) and middle-aged (mean=3.519, p<<.01) adults. Edoardo received higher suitability scores by seniors (mean=4.333) rather than adolescents (mean=3.283, p<<.01), young (mean=2.851, p<<.01) and middle-aged (mean=2.494, p<<.01) adults. Clara received higher suitability scores by seniors (mean=4.401) rather than adolescents (mean=3.298, p<<.01), young adults (mean=2.981, p<<.01) and middle-aged participants (mean=2.639, p<<.01). Edoardo received higher suitability scores by adolescents (mean=3.283) rather than middle-aged (mean=2.494, p=.006), and Clara received higher suitability scores by adolescents (mean=3.298) rather than middle-aged participants (mean=2.639, p=.041) in performing healthcare occupations.

b) Concerning healthcare scores, Giulia (mean=3.608) received by young adults significantly higher scores than Edoardo (mean=2.851, p<<.01) and Clara (mean=2.981, p=.015). Giulia (mean=3.519) received by middle-aged adults significantly higher scores than Edoardo (mean=2.494, p<<.01) and Clara (mean=2.639, p<<.01). Antonio (mean=3.205) received by middle-aged adults significantly higher scores than Edoardo (mean=2.494, p=.001) and Clara (mean=2.639, p=.014).

*Housework*

No participants' gender effects (F (1,178) = 3.590, p=.060) were observed. Significant differences among age groups (F (3,178) = 17.082, p<<.01) emerged. Bonferroni's post hoc tests revealed that seniors (mean=4.020) considered the synthetic voices significantly more suitable than adolescents (mean=3.090, p<<.01), young adults (mean=3.345, p=.002) and middle-aged participants (mean=2.771, p<<.01) in performing housework tasks. Additionally, young adults (mean=3.345) considered the proposed voices more appropriate than middle-aged participants (mean=2.771, p=.013) to housework occupations.

A significant interaction emerged between participants' gender and age groups (F (3,178) = 4.788, p=.003). Bonferroni's post hoc tests were performed for each single factor (age groups and participants' gender). Tests showed that:

a) Concerning the age groups, male seniors (mean=4.511) attributed higher suitability scores to the voices than male adolescents (mean=3.073, p<<.01), male young (mean=3.190, p<<.01) and male middle-aged adults (mean=2.943, p<<.01) in performing housework occupations. Female middle-aged participants (mean=2.598) attributed lower suitability scores to the voices than female young (mean=3.500, p=.003) and female (mean=3.529, p=.002) older adults.

b) Concerning participants' gender, male older (mean=4.511) considered the voices as more suited than female older (mean=3.529, p<<.01) adults in performing housework tasks.

The four proposed voices (F (3,534) = 10.401, p<<.01) differed significantly in their suitability scores to perform housework. Bonferroni's post hoc tests revealed that Edoardo (mean=3.068) was considered significantly less suitable than Antonio (mean=3.326, p=.012), Giulia (mean=3.522, p<<.01) and Clara (mean=3.310, p=.025) for housework.

A significant interaction emerged between age groups and the voices' suitability scores in performing housework (F (9,534) = 2.390, p=.012). Bonferroni's post hoc tests were performed for each single factor (age groups and housework scores). Tests revealed that:

a) Concerning age groups, Antonio was more appreciated by seniors (mean=4.008) rather than adolescents (mean=3.152, p=.001), young (mean=3.336, p=0.20) and middle-aged adults (mean=2.806, p<<.01). Seniors (mean= 4.030) attributed higher suitability scores than adolescents (mean=3.280, p=.006) and middle-aged participants (mean=3.294, p=.007) to Giulia in performing housework. Seniors (mean=4.008) rated Edoardo more suited than adolescents (mean=2.884, p<<.01), young (mean=3.106, p=.001) and middle-aged (mean=2.274, p<<.01) adults in performing housework. Young adults (mean=3.106) rated Edoardo more suitable than middle-aged (mean=2.274, p=.004) adults. Seniors (mean=4.033) rated Clara more appropriate than adolescents (mean=3.045, p<<.01) and middle-aged participants (mean=2.708, p<<.01). Young adults (mean=3.455) rated Clara more suited than middle-aged participants (mean=2.708, p=.010) in performing housework.

b) Concerning housework suitability scores, middle-aged participants rated Giulia (mean=3.294) more appropriate than Antonio (mean=2.806, p=.014), Edoardo (mean=2.274, p<<.01) and Clara (mean=2.708, p=.004), and Antonio (mean=2.806) more suitable than Edoardo (mean=2.274, p=.009) in performing housework.

*Protection and security tasks*

No participants' gender effect (F (1,178) = .010, p=.922) emerged. Significant differences among age groups (F (3,178) =8.384, p<<.01) were found. Bonferroni's post hoc tests revealed that seniors (mean=3.738) considered the voices significantly more suitable than adolescents (mean=3.156, p=.012) and middle-aged (mean=2.821, p<<.01) adults in performing protections and security tasks. Suitability scores for protection and security tasks differed significantly among the four proposed voices (F (3,534) = 5.617,

p=.001). Bonferroni's post hoc tests revealed that Antonio (mean= 3.436) was considered significantly more qualified than Edoardo (mean= 3.112, p=.011) and Clara (mean=3.118, p=.005) for protection and security tasks.

A significant interaction emerged between age groups and suitability scores for protection and security tasks (F (9,534) = 2.827, p=.003). Bonferroni's post hoc tests were performed for each single factor (age groups and suitability scores for protection and security tasks).

Tests revealed that:

a) Concerning age groups, seniors (mean=3.759) considered Antonio significantly more qualified than adolescents (mean=3.086, p=.035) in performing these specific tasks. Seniors (mean=3.716) also considered Edoardo as more suited than middle-aged (mean=2.513, p<<.01) participants. Seniors (mean=3.805) considered Clara as more suitable than adolescents (mean=3.113, p=.034), young (mean=3.131, p=.039) and middle-aged (mean=2.422, p<<.01) participants. Middle-aged participants (mean=2.422) considered Clara significantly less qualified for protection and security tasks than adolescents (mean=3.113, p=.041) and young adults (mean=3.131, p=.031).

b) Concerning suitability scores for protection and security tasks, young adults attributed to Antonio (mean=3.657) higher suitability scores than those attributed to Giulia (mean=3.164, p=.033) and Clara (mean=3.131, p=.030). Middle-aged adults attributed to Antonio (mean=3.241) higher suitability scores than Edoardo (mean=2.513, p=.003) and Clara (mean=2.422, p<<.01), as well as considered Giulia (mean=3.110) which was more appropriate than Edoardo (mean=2.513, p=.010) and Clara (mean=2.422, p=.006).

*Front office*

No participants' gender effect (F (1,178) = .000, p=.989) and no significant differences among age groups (F (3,178) = .208, p=.891) were observed. Suitability to front office tasks was rated significantly different among the four proposed voices (F (3,534) = 14.335, p<<.01). Bonferroni's post hoc tests revealed that Antonio (mean=3.317) was considered significantly more suited than Edoardo (mean=2,931, p<<.01) and Clara (mean=2.916, p<<.01), and Giulia (mean=3.427) was considered significantly more suitable than Edoardo (mean=2,931, p<<.01) and Clara (mean=2.916, p<<.01) for front office tasks.

A significant interaction emerged between age groups and suitability scores to perform front office tasks (F (9,534) = 4.601, p<<.01). Bonferroni's post hoc tests were performed for each single factor (age groups and front office tasks). Tests revealed that:

a) Concerning age groups, middle-aged participants (mean=3.759) considered Giulia more qualified than seniors (mean=3.034, p=.034) in performing front office tasks.

b) Concerning voices' suitability scores for front office tasks, young adults attributed to Giulia (mean=3.636) higher suitability scores than Edoardo (mean=2.755, p<<.01) and Clara (mean=2.967, p=.010), and to Antonio (mean=3.532) higher suitability scores than Edoardo (mean=2.755, p<<.01) and Clara (mean=2.967, p=.028). Middle-aged adults, attributed to Giulia (mean=3.759) higher suitability scores than Edoardo (mean=2.714, p<<.01) and Clara (mean=2.643, p<<.01), and to Antonio (mean=3.559) was higher suitability scores than Edoardo (mean=2.714, p<<.01) and Clara (mean=2.643, p<<.01).